\documentclass[reprint,amsmath,amssymb,aps,prb,superscriptaddress,twocolumn]{revtex4-2}
\usepackage{graphicx}
\usepackage{graphics}
\usepackage{dcolumn}
\usepackage{bm}
\usepackage{amsfonts}
\usepackage{amssymb}
\usepackage{xcolor}
\usepackage{multirow}
\usepackage{mathtools}
\usepackage[tight]{subfigure}
\usepackage{relsize}

\definecolor{green}{rgb}{0,0.6,0.1}

\begin{document}

\title{Spin-fluctuation-mediated unconventional $d_{xy}$-wave superconductivity in ThFeAsN from first-principles}

\author{Guang-Yu Guo}
\email{gyguo@phys.ntu.edu.tw}
\affiliation{Department of Physics and Center for Theoretical Physics, National Taiwan University, Taipei 10617, Taiwan\looseness=-1}
\affiliation{Physics Division, National Center for Theoretical Sciences, Taipei 10617, Taiwan\looseness=-1}

\author{Jau-Wen Liu}
\affiliation{Department of Physics, National Taiwan University, Taipei 10617, Taiwan\looseness=-1}

\author{Mitsuaki Kawamura}
\affiliation{Division of Intelligent System Engineering, Yokohama National University, Yokohama 240-8501, Japan\looseness=-1}

\date{\today}

\begin{abstract}
Superconducting (SC) pairing mechanism, origin of high $T_c$ and symmetry of SC order parameter 
in Fe-based superconductors are among the important unsolved problems in condensed matter and materials physics.
We study the SC properties of ThFeAsN, a Fe-based high $T_c$ superconductor,
by {\it ab initio} superconducting density functional theory calculations 
with electron-phonon coupling, screened static and dynamic electron-electron Coulomb repulsion 
and spin fluctuation (SF) mediated pair-interaction fully taken into account. 
Our calculations reveal that ThFeAsN is a SF-mediated multiband superconductor 
with the calculated $T_c$ of 22.4 K and the $d_{xy}$-wave SC order parameter 
with different signs on different Fermi surface sheets, 
in consistent with most experiments.
We also present distinct SC properties such as quasiparticle density of states 
and ultrasonic attenuation coefficient which can be immediately verified by experiments.
\end{abstract}

\maketitle
\begin{figure}[!htb]
\centering
\includegraphics[width=0.90\columnwidth]{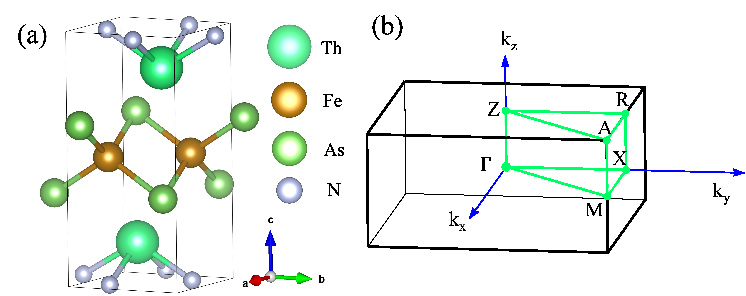}
\includegraphics[width=0.90\columnwidth]{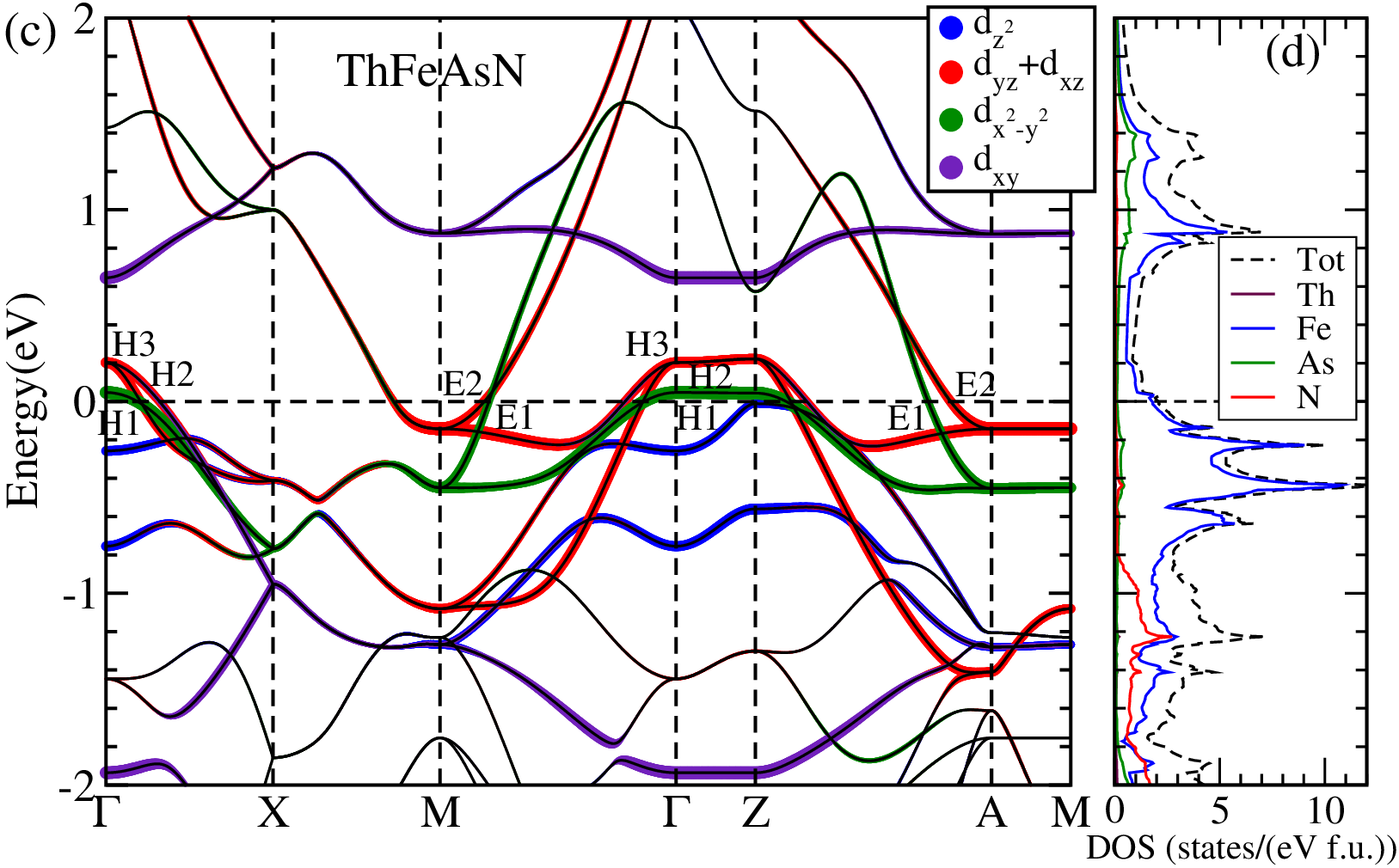}
\caption{(a) Crystal structure and (b) Brillouin zone of ThFeAsN.
(c) Fe $d$ orbital-projected band structure and (d) total and atom-decomposed DOS of ThFeAsN.
In (c), the Fe $d$ orbital weights are proportional to the line widths.}
\label{fig:crystal-bands}
\end{figure}

Since the discovery of high critical temperature ($T_c$) superconductivity in LaFeAs(O,F) in 2008~\cite{Kamihara2008}, 
many Fe-based superconductors (FeSCs) have been found and lots of experimental and theoretical investigations 
have been carried out on FeSCs~\cite{Stewart2011,Chen2014,Shibauchi2020}. 
However, as for high $T_c$ cuprate superconductors, the precise mechanism of the superconducting (SC) pairing, 
the origin of high $T_c$ and the SC gap structure in FeSCs 
remain poorly understood~\cite{Stewart2011,Chen2014,Shibauchi2020}.
Parent FeSCs are generally antiferromagnetic (AF) and metallic but nonsuperconducting.
When chemically doped, the magnetic order is suppressed
and the superconductivity emerges. Nevertheless, the conventional $s$-wave phonon-mediated 
Bardeen-Cooper-Schrieffer
(BCS) mechanism~\cite{BCS1957} was ruled out because electron-phonon
coupling (EPC) was found too weak to explain the high $T_c$~\cite{Boeri2008,Sen2020,Schrodi2021}.
Apart from their SC property, many FeSCs also exhibit intriguing normal
state properties such as orbital density wave, structural phase transition,
and nematic order~\cite{Dong2008,Stewart2011,Chen2014,Fernandes2014,Shibauchi2020},
which are distinctly different from high-$T_c$ cuprates.
Almost all FeSCs have a tetragonal structure with FeAs/FeSe layers
as the basic building blocks [Fig.~\ref{fig:crystal-bands}(a)].
Also, FeSCs have Fe $d$-orbital dominated bands
crossing the Fermi level~\cite{Stewart2011,Chen2014,Shibauchi2020} 
[see also Fig.~\ref{fig:crystal-bands}(c)], and the Fermi surface (FS) 
consists of several electron and hole FS sheets at the Brillouin zone (BZ) 
corners and the BZ center [Fig.~\ref{fig:crystal-bands}(b)], respectively. 
Moreover, pronounced AF spin fluctuation (SF) was observed in the doped systems, 
although long-range magnetic order was generally suppressed.
Based on the multiband FS topology and the strong SF in the normal state, 
a sign-changing $s_{\pm}$-wave pairing on the disconnected hole and electron 
FS sheets was proposed~\cite{Mazin2008,Kuroki2008}. This nodeless $s_{\pm}$-wave pairing scenario 
is supported by, e.g., inelastic neutron scattering~\cite{Christianson2008} and 
quasiparticle heat transport~\cite{Ding2009} measurements.
Nevertheless, a nodal $d$-wave SC state was also proposed for
heavily hole-doped K$_x$Ba$_{1-x}$Fe$_2$As$_2$~\cite{Thomale2011}.
Subsequent angle-resolved photoemission spectroscopy (ARPES) measurements~\cite{Okazaki2012}
revealed that KFe$_2$As$_2$ has the $s$-wave symmetry overall with a nodeless gap on the innermost hole 
pocket, an octet line node structure on the middle cylinder and a small gap on the outer one.
Recent muon spin rotation ($\mu$SR) measurements reported evidence for
a nodal multigap superconductivity in KCa$_2$Fe$_4$As$_4$F$_2$~\cite{Smidman2018}.
Thus, the superconductivity in FeSCs is still 
not fully understood~\cite{Stewart2011,Chen2014,Shibauchi2020}. 

Among FeSCs, stoichiometric ThFeAsN with a high $T_c$ of
29 K~\cite{Wang2016}, is an ideal system for studying the superconductivity in FeSCs. 
Unlike most other FeSCs, ThFeAsN does not show long-range magnetic 
order~\cite{Albedah2017,Mao2017,Shiroka2017,Adroja2017}, although  
significant magnetic fluctuations above 35 K was observed in $\mu$SR
and nuclear magnetic resonance (NMR) experiments~\cite{Shiroka2017,Adroja2017}. 
Thus, ThFeAsN is an intrinsic superconductor without chemical doping, 
which inevitably introduces randomness
and can even change the crystal structure. Moreover, unlike FeSe~\cite{Shibauchi2020},
ThFeAsN does not exhibit structural phase transition nor nematic order.
Recent multichannel Eliashberg equation calculations with input from {\it ab initio} calculations
indicated that the AF SF would play a dominant
role and ThFeAsN would have the $s_{\pm}$-wave SC gaps. 
However, the predicted $T_c$ is 3.15 K~\cite{Schrodi2021}.
Furthermore, when the EPC and charge fluctuation are also accounted for, 
no superconductivity was found down to 2 K.~\cite{Schrodi2021}

Here we study the SC properties of ThFeAsN by performing {\it ab initio} superconducting 
density functional theory (SCDFT)~\cite{Oliveira1988,Lueders2005,Marques2005} calculations
with the EPC-induced SC pairing, screened static and dynamic electron-electron 
(e-e) Coulomb repulsion and SF-mediated pair-interaction being fully accounted for. 
Our SCDFT calculations predict
that ThFeAsN is an unconventional superconductor with an even-parity SC order parameter having
the $B_{2g}$ symmetry. Its SC gap structure consists of multiband quasi-two-dimensional $d_{xy}$-waves
with FS sheet-dependent signs. The calculated $T_c$ of 22.4 K agrees quite well
with the experimental value of 29 K. 

{\it Theory and method}. 
In the SCDFT, in addition to the EPC, the screened e-e Coulomb repulsion and 
SF-mediated pairing interaction are also treated in a first-principles manner 
(see \cite{Lueders2005,Marques2005,Essenberger2014,Kawamura2020} and references therein).
Using the normal state properties from {\it ab initio} calculations as inputs, 
one solves the SC gap equation 
\begin{align}
\Delta_{n\bf{k}} = - \frac{1}{2}\sum_{m{\bf k}'}\frac{K_{n{\bf k}m{\bf k}'}(\xi_{n{\bf k}},\xi_{m{\bf k}'})}{1+Z_{n{\bf k}}(\xi_{n{\bf k}})}\frac{\mathrm{tanh}[E_{m{\bf k}'}/2k_BT]}{E_{m{\bf k}'}}\Delta_{m{\bf k}'},
\label{eq:gap-eq}
\end{align}
where $\Delta_{n{\bf k}}$ is the gap function, $n$ and $\bf k$ denote
the band index and the crystal momentum, respectively. 
Also, $E_{n\bf{k}} = \sqrt{\xi_{n{\bf k}}^2+|\Delta_{n{\bf k}}|^2}$
and $\xi_{n{\bf k}} = \varepsilon_{n{\bf k}} - \mu$ which is the DFT eigen-energy ($\varepsilon_{n{\bf k}}$) 
measured from chemical potential $\mu$.
The integration kernels $K_{n{\bf k}m{\bf k}'}(\xi_{n{\bf k}},\xi_{m{\bf k}'})$ include the SC-pair
breaking and creating interactions and comprises the EPC, the e-e Coulomb repulsion and the SF kernel,
i.e., $K_{n{\bf k}m{\bf k}'}(\xi_{n{\bf k}},\xi_{m{\bf k}'}) = K^{ep}_{n{\bf k}m{\bf k}'}(\xi_{n{\bf k}},\xi_{m{\bf k}'}) 
+ K^{ee}_{n{\bf k}m{\bf k}'}(\xi_{n{\bf k}},\xi_{m{\bf k}'}) + K^{SF}_{n{\bf k}m{\bf k}'}(\xi_{n{\bf k}},\xi_{m{\bf k}'})$. 
The renormalization $Z_{n{\bf k}}(\xi_{n{\bf k}})$ contains only the EPC and SF terms, i.e.,
$Z_{n{\bf k}}(\xi_{n{\bf k}}) = Z_{n{\bf k}}^{ep}(\xi_{n{\bf k}}) + Z_{n{\bf k}}^{SF}(\xi_{n{\bf k}})$, 
because the e-e Coulomb repulsion is already included in the DFT eigenvalues $\xi_{n{\bf k}}$.
The expressions for kernel $K_{n{\bf k}m{\bf k}'}(\xi_{n{\bf k}},\xi_{m{\bf k}'})$ and renormalization
$Z_{n{\bf k}}(\xi_{n{\bf k}})$ have already been given in, e.g., ~\cite{Lueders2005,Marques2005,Essenberger2014,Kawamura2020}.
The crystal structure of ThFeAsN and computational details are given in Appendix A.

{\it Electronic structure and Fermi surface}. The calculated electronic energy bands and 
density of states (DOS) of ThFeAsN are displayed in 
Figs.~\ref{fig:crystal-bands}(c) and ~\ref{fig:crystal-bands}(d), respectively. 
Figure ~\ref{fig:crystal-bands}(c) shows that five bands (labelled H1, H2, H3, and E1 and E2) cross the Fermi level ($E_F$).
As a result, the FS consists of five cylinder-like sheets 
(Fig. ~\ref{fig:gap_FS}), indicating a two-dimension (2D)-like band structure due to
the  weak interaction between the Fe-As layers [see Fig. 1(a)].
The H1, H2, H3 bands form three cylindrical hole sheets centered on $\Gamma$, 
while the E1 and E2 bands form two cylindrical electron sheets located at M 
[see Figs.~\ref{fig:crystal-bands}(c) and ~\ref{fig:gap_FS}].
The DOS near the $E_F$ is dominated by Fe $d$-orbitals [see Fig.~\ref{fig:crystal-bands}(d)].
The total DOS at the $E_F$ is 2.066 states/eV/f.u., and atom-decomposed DOSs at the $E_F$
are 0.010, 1.780, 0.109 and 0.030 states/eV/f.u., respectively, for Th, Fe, As, and N atoms.
Figure~\ref{fig:crystal-bands}(c) indicates that the H2 and H3 bands
along the $\Gamma$-Z line as well as the E2 pocket along the M-A line arise mainly from Fe $d_{xz}+d_{yz}$ 
orbitals while the H1 band is dominated by Fe $d_{x^2-y^2}$ orbital [see Fig.~\ref{fig:crystal-bands}(c)]. 
The E1 band is also dominated by Fe $d_{x^2-y^2}$ orbital [Fig.~\ref{fig:crystal-bands}(c)].

\begin{figure}[!htb]
\centering
\includegraphics[width=0.85\columnwidth]{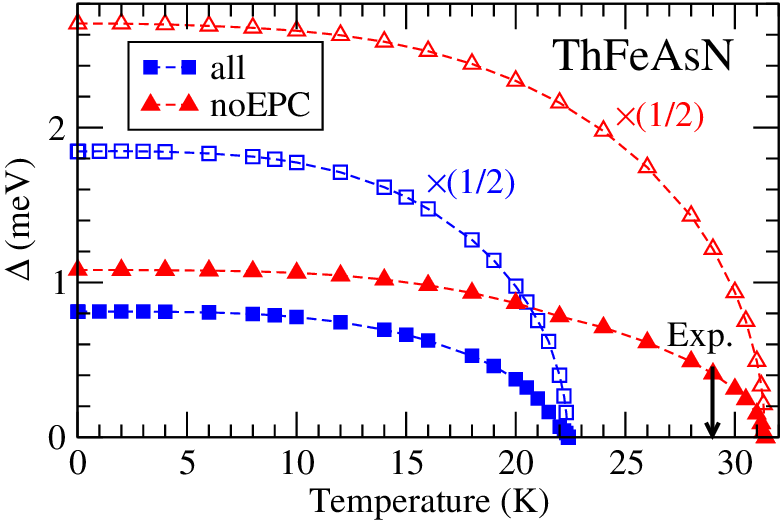}
\caption{Averaged (solid symbols) and maximum (open symbols) SC gap $\Delta$ 
vs. temperature $T$ from the full (squares) and no EPC (up-triangles) SCDFT calculations.
}
\label{fig:gap_T}
\end{figure}

\begin{table}[b]
\caption{\label{tab:SC-property}%
Calculated superconducting (SC) critical temperature ($T_c$),  averaged ($\Delta_0$) and maximum ($\Delta_m$) 
SC gap at $T = 0$ K, EPC constant $\lambda$, logarithmical average phonon frequency ($\omega_{ln}$),
screened Coulomb potential ($\mu$), and renormalization factor ($Z$)
for (a) EPC only, (b) EPC plus Coulomb pseudopotential ($\mu$)
(c) EPC plus $\mu$ and SF, and (d) no EPC. Calculated DOS at the Fermi level
$N_F = 2.066$ states/eV/f.u.}
\begin{ruledtabular}
\begin{tabular}{ccccc}
                    &(a) EPC & (b) EPC+$\mu$ &(c) EPC+$\mu$+SF &(d) $\mu$+SF \\
\hline
$T_{c}$ (K)         & 0.17     & 0.0   & 22.4 (29$^a$) & 31.4 \\
$\Delta_0$ (meV)    & 0.024    & 0.0   & 0.811 & 1.081 \\
$\Delta_m$ (meV)    & 0.037    & 0.0   & 3.691 & 5.344 \\
$\omega_{ln}$ (K)   & 213      & 213   & 213   & 0   \\
$\lambda$           & 0.127    & 0.127 & 0.127 & 0.0 \\
$\mu$               & 0.0      & 0.687 & 1.304 & 1.304 \\
$Z$                 & 0.262    & 0.741 & 1.102 & 0.843 \\
\end{tabular}
\end{ruledtabular}
$^a$The experimental $T_c$ from Ref.~\cite{Wang2016}.
\end{table}

\begin{figure}[!htb]
\centering
\includegraphics[width=0.85\columnwidth]{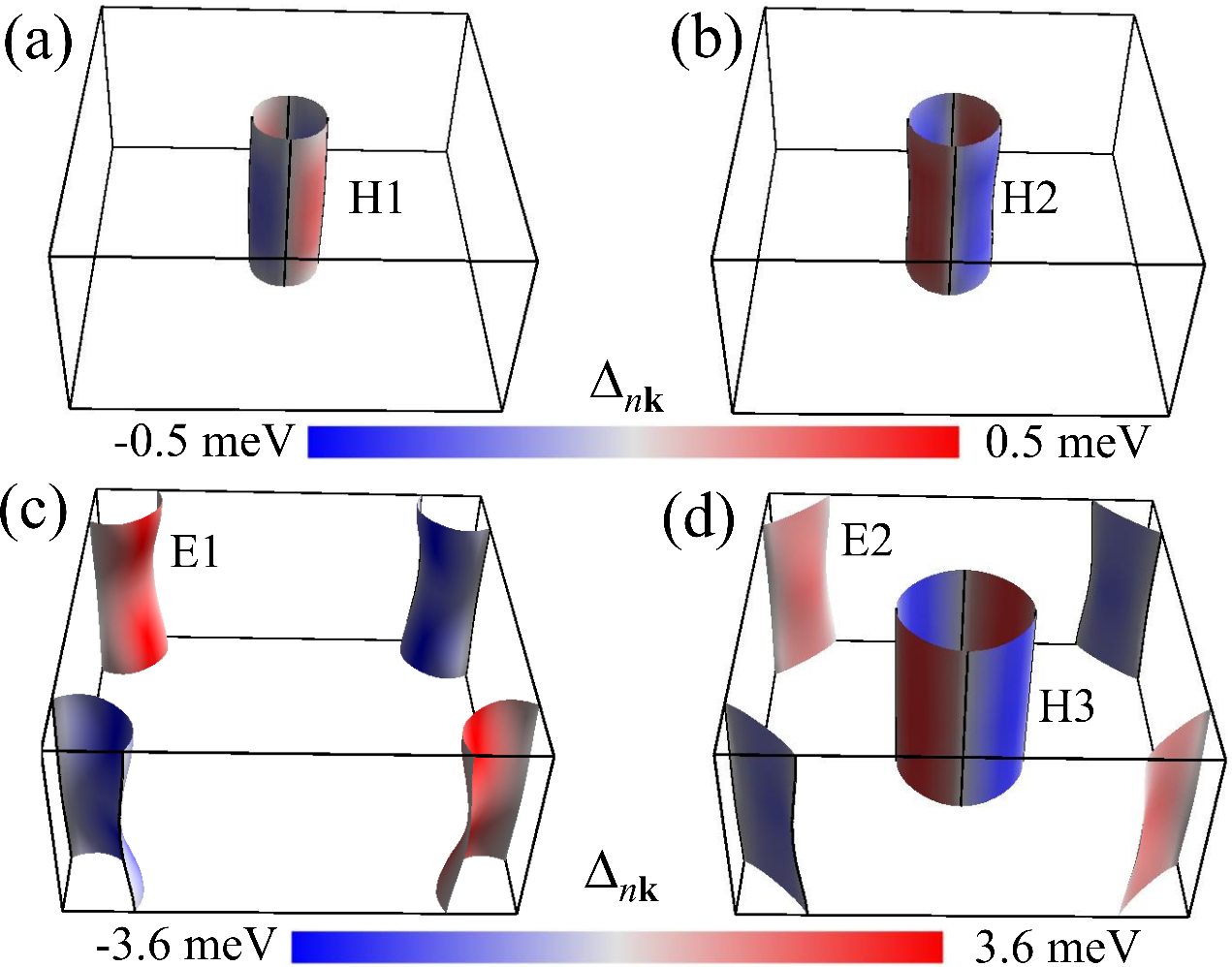}
\caption{Momentum ${\bf{k}}$-dependent SC gap $\Delta_{n{\bf k}}$ on the Fermi surface at 0 K: 
hole pockets H1 (a) and H2(b), electron pocket E1 (c), hole pocket H3 and electron pocket E2 (d).
The black vertical lines denote the nodal lines.} 
\label{fig:gap_FS}
\end{figure}

{\it Transition temperature and gap symmetry}. 
To determine the transition temperature $T_c$, 
we solve the SC gap equation [Eq.~(\ref{eq:gap-eq})] at many temperatures. 
In Fig.~\ref{fig:gap_T}, the averaged and maximum SC gaps are plotted as a function of $T$. 
From Fig.~\ref{fig:gap_T}, we obtain the averaged $\Delta(0)=0.811$ meV at $T = 0$ K 
and $T_c = 22.4$ K, as listed in Table I. The maximum SC gap $\Delta(0)$ is 3.691 meV.
The calculated $T_c$ agrees quite well with the experimental value of 29 K~\cite{Wang2016}.
To study the microscopic mechanism, we do the following "computer experiments". 
First, we switch-off both the SF and screened e-e Coulomb repulsion 
(i.e., the case of purely phonon-mediated e-e pairing mechanism). 
We find that $T_c$ reduces to merely 0.17 K. ThFeAsN becomes a conventional anisotropic $s$-wave
superconductor with the SC gap varying from 0.017 to 0.040 meV. This is similar to
the case of $\gamma$-BiPd, which is an anisotropic $s$-wave superconductor~\cite{Keshri2025}. 
Second, if we turn-on the screened e-e repulsion, the weak phonon-mediated 
superconductivity is completely suppressed (Table I).
Finally, let us keep  the SF interaction and screened Coulomb repulsion but switch-off the EPC. 
Remarkably, the calculated $T_c$ goes up to 31.4 K (see Table I and Fig. 2),
slightly exceeding the experimental $T_c$ of 29 K~\cite{Wang2016}. 
These "computer experiments" clearly demonstrate that the superconductivity in ThFeAsN 
is unconventional and arises from the SF-mediated e-e pairing.

Now let us examine the obtained SC gap function.
Band and momentum $n${\bf k}-dependent gap function $\Delta_{n{\bf k}}$
on the FS at $ T = 0$ K is displayed in Figs.~\ref{fig:gap_FS}.
Figure ~\ref{fig:gap_FS} shows that ThFeAsN exhibits 
quasi-two-dimensional $d_{xy}$-waves with different signs on different FS sheets. 
Specifically, all the gap functions on the five FS sheets 
are $d_{xy}$-waves with the $B_{2g}$ symmetry (see Appendix B and Table II).
In particular, the gap function $\Delta_{n{\bf k}}$ on the largest hole pocket (H3)
at the BZ center forms a $d_{xy}$-wave with vertical nodal lines in the [100] and [010] direction
on the almost perfect cylinder [Fig. ~\ref{fig:gap_FS}(d)]. 
Figure ~\ref{fig:gap_FS}(b) shows that the $\Delta_{n{\bf k}}$ on the second largest hole 
pocket (H2) is also a $d_{xy}$-wave with the same sign as that on the H3 pocket.
We notice that this gap structure with vertical nodal lines is similar to the octet-line node SC gap structure
on the cylindrical FS sheet in KFe$_2$As$_2$ revealed by ARPES~\cite{Okazaki2012}.
Interestingly, the gap functions on the two electron pockets E1 and E2 at
the BZ corners are also $d_{xy}$-waves, albeit with signs being 
opposite to that on the H2 and H3 pockets. 
Therefore, the SC gap structure of ThFeAsN could be considered
as the $d_{xy}({\pm}$)-waves, being similar to 
the $s_{\pm}$-waves previously proposed for LaFeAsO$_{1-x}$F$_x$~\cite{Mazin2008,Kuroki2008}.
Nevertheless, there is a notable difference, i.e., the $s_{\pm}$-waves have no vertical line nodes.
Finally, the smallest hole cylinder (H1) at the BZ center too hosts a $d_{xy}$-wave
with a sign opposite to that on the H2 and H3 pockets [see Figs. ~\ref{fig:gap_FS}(a), ~\ref{fig:gap_FS}(b)
and ~\ref{fig:gap_FS}(d)].

Limited experimental studies on the superconductivity in ThFeAsN have given somewhat inconsistent results
(see, e.g., ~\cite{Shiroka2017,Adroja2017}).
This is perhaps natural, given the sophisticated nature of the FS topology with five bands crossing 
the Fermi level [Fig. 1(c) and Fig. ~\ref{fig:gap_FS}]. $\mu$SR and NMR experiments performed 
by Shiroka {\it et al.}~\cite{Shiroka2017} showed that ThFeAsN exhibits strong SF  
above $\sim$35 K without magnetic order. These measurements~\cite{Shiroka2017} also indicated 
that the electronic excitation 
spectrum in the SC state would be best explained by a two-gap nodeless $s$-wave model. 
Adroja {\it et al.} also carried out transverse field $\mu$SR measurements and, on the other hand, 
found that their temperature and field dependent results could be better fitted with two-gap models, 
than either the single-gap isotropic $s$-wave or $d$-wave model~\cite{Adroja2017}. They also noted that 
similar good fits could be obtained with either isotropic ($s + s$)-wave or an ($s+d$)-wave model with line nodes. 
Furthermore, their field-dependent specific heat measurements indicated a nodal SC gap~\cite{Adroja2017},
suggesting that the ($s + d$)-wave model best explains the gap structure of ThFeAsN.
These later results are consistent with our finding of multiband superconductivity 
with sign changing $d_{xy}$-waves on different FS sheets with vertical nodal lines (Fig.~\ref{fig:gap_FS}).
In light of our predicted SC gap structure, it would be
helpful to examine again the data of these $\mu$SR measurements~\cite{Adroja2017,Shiroka2017},
in order to resolve these different interpretations.
High resolution ARPES measurements, being capable of probing the band-dependent SC gap 
structure~\cite{Okazaki2012,Mou2011}, on ThFeAsN in the SC state will be highly desirable. 

\begin{figure}[!htb]
\centering
\includegraphics[width=0.80\columnwidth]{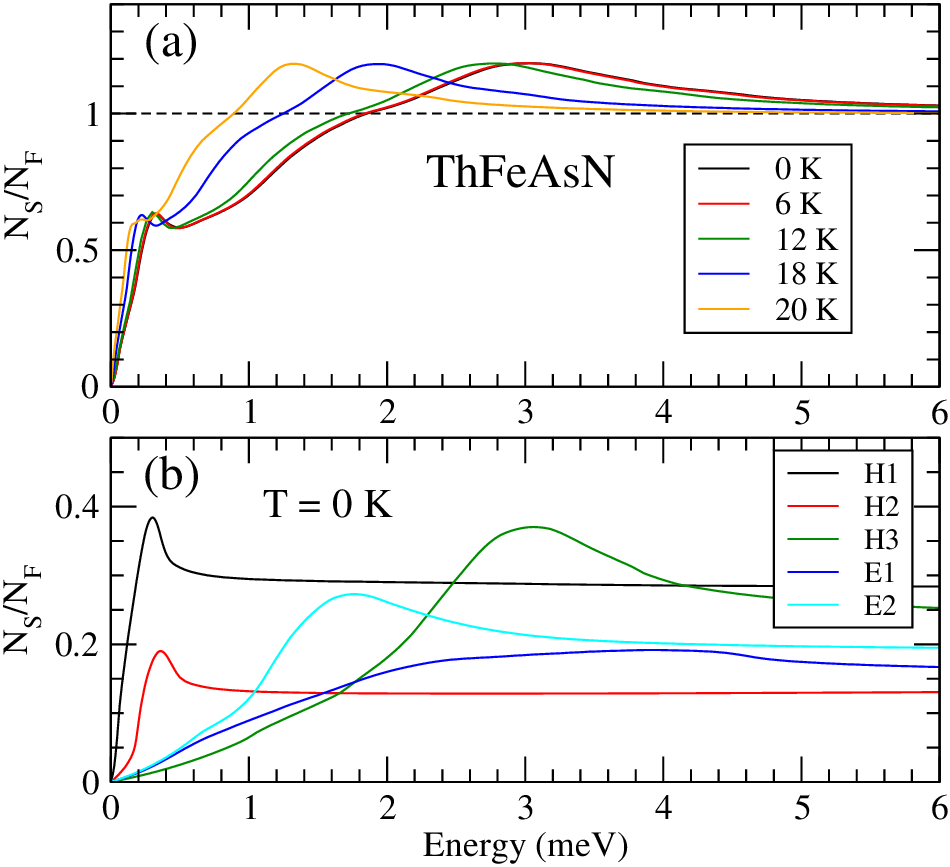}
\caption{(a) Normalized SC quasiparticle DOS $N_S/N_F$ vs. energy for several temperatures. 
(b) FS sheet-decomposed normalized $N_S/N_F$ at $T = 0$ K.}
\label{fig:qpdos}
\end{figure}

{\it Quasiparticle density of states}. 
In order to stimulate further experiments such as scanning tunneling spectroscopy 
(STS) \cite{Fischer2007} and ultrasonic attenuation~\cite{Watanabe2004} on the SC properties 
of ThFeAsN, we also calculate the $T$-dependent SC quasiparticle DOS (QPDOS) [$N_s(T)$] spectrum and 
ultrasonic attenuation coefficient using the obtained gap function $\Delta_{n\bf{k}}$.
We depict the normalized total and band-resolved SC QPDOS ($N_s/N_F$) spectra 
in Figs.~\ref{fig:qpdos}(a) and ~\ref{fig:qpdos}(b), respectively. 
Because nodeless and nodal SC gap functions exhibit distinctly different features 
in the QPDOS spectrum, STS experiments would allow us to study the SC gap structure 
such as the gap symmetry and gap size distribution.~\cite{Fischer2007}
Figure~\ref{fig:qpdos}(a) indicates that the profiles of all the QPDOS spectra 
are almost the same except that the energy positions of the spectral
features move towards the zero energy as $T$ is lowered.
Thus, for clarity, we focus on the QPDOS spectrum at $T = 0$ K. 
Figure ~\ref{fig:qpdos}(a) shows that the $N_s/N_F$(0) spectrum increases linearly with energy
from 0 to $\sim$0.3 meV, where it decreases slightly until $\sim$0.5 meV.
Then it increases monotonically (although nonlinearly) and reaches its maximum at $\sim$3.0 meV,
where it starts decreasing slowly until it reaches nearly 1.0 beyond 6.0 meV.
Clearly, the QPDOS spectrum has a V-shape below $\sim$0.3 meV [Fig. ~\ref{fig:qpdos}(a)],
consistent with the nodal $d$-wave superconductivity predicted for ThFeAsN.
To understand the origins of these interesting features in the QPDOS spectrum, we examine 
the band-resolved $N_s/N_F$ spectra at $T = 0$ K in Fig.~\ref{fig:qpdos}(b).
Figure~\ref{fig:qpdos}(b) suggests that all the $N_s/N_F$ spectra initially increases almost linearly,
forming a V-shape function caused by the nodal $d_{xy}$-wave gap function on
their respective FS pocket. The $N_s/N_F$ spectra then continue increasing monotonically
until they reaches their respective maximum, which depends on their SC gap maximum.
For example, hole pockets H1 and H2 have small SC gaps 
of $\sim$0.3 meV [Figs.~\ref{fig:gap_FS}(a) and ~\ref{fig:gap_FS}(b)],
leading to the peak at $\sim$0.3 meV in both the total [Fig.~\ref{fig:qpdos}(a)] and
band-resolved  [Fig.~\ref{fig:qpdos}(b)] $N_s/N_F$ spectra. Meanwhile,
hole pocket H3 has the largest SC gap, giving rise to the broad peak at $\sim$3.0 meV  
in both  [Fig.~\ref{fig:qpdos}(a)] and band-resolved  [Fig.~\ref{fig:qpdos}(b)] $N_s/N_F$ spectra.

\begin{figure}[!htb]
\centering
\includegraphics[width=0.85\columnwidth]{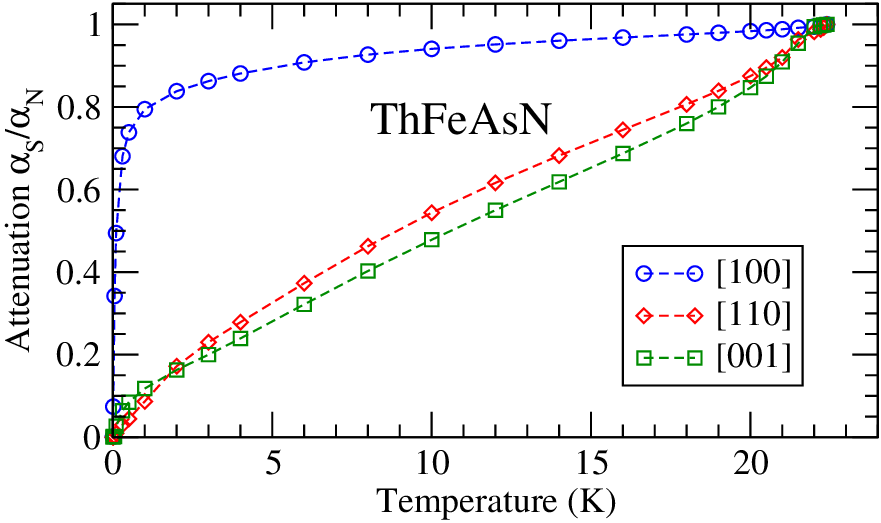}
\caption{Normalized longitudinal ultrasonic attenuation coefficients $\alpha_S/\alpha_N(T)$ where $\alpha_S$ ($\alpha_N$)
is the ultrasonic attenuation coefficient in the superconducting (normal) state.
}
\label{fig:attenuation}
\end{figure}

{\it Ultrasonic attenuation}. 
Ultrasonic attenuation $\alpha (T)$ with its directional property is another powerful probe  of 
SC gap symmetry and positions of nodal points on the FS.~\cite{Watanabe2004,Vekhter1999}
In Fig.~\ref{fig:attenuation}, we display normalized longitudinal ultrasonic attenuation 
coefficients $\alpha_S/\alpha_N$ along three high symmetry directions calculated  
using the obtained $\Delta_{n{\bf k}}$ and Eq.~(\ref{eq:ultrasonic}) in Appendix C. 
First, we notice that the $\alpha_S/\alpha_N (T)$ is highly direction-dependent.
In particular, the ultrasonic attenuation along the [100] direction is much stronger 
and decreases much slowly (above $\sim$2.0 K) with decreasing $T$ than in the [110] and [001] directions 
(Fig.~\ref{fig:attenuation}). This is caused by the presence of the vertical nodal lines in the [100]
direction in the $d_{xy}$ gap function on all the five FS sheets (see Fig. ~\ref{fig:gap_FS}),
because a longitudinal sound wave couples dominantly with the quasiparticles 
having momentum parallel to sound propagation direction~\cite{Watanabe2004}.
Second, at low temperatures, $\alpha_S/\alpha_N (T)$ in a fully gap $s$-wave superconductor 
such as Sn increases with $T$ exponentially~\cite{Morse1957}. 
In contrast, Fig. ~\ref{fig:attenuation} shows that in the very low temperature region (e.g., $T < 1.0$ K), 
$\alpha_S/\alpha_N (T)$ in the [100] and [110] directions increases nearly linearly with $T$ and
that along the [001] direction increases with $T$ in a non-exponential manner, consistent
with the nodal $d_{xy}$-wave superconductivity~\cite{Vekhter1999} in ThFeAsN discovered in this work.
We believe that this prediction of unconventional $\alpha_S/\alpha_N (T)$
will stimulate ultrasonic attenuation experiments on ThFeAsN in the near future.

\begin{figure}[!htb]
\centering
\includegraphics[width=0.70\columnwidth]{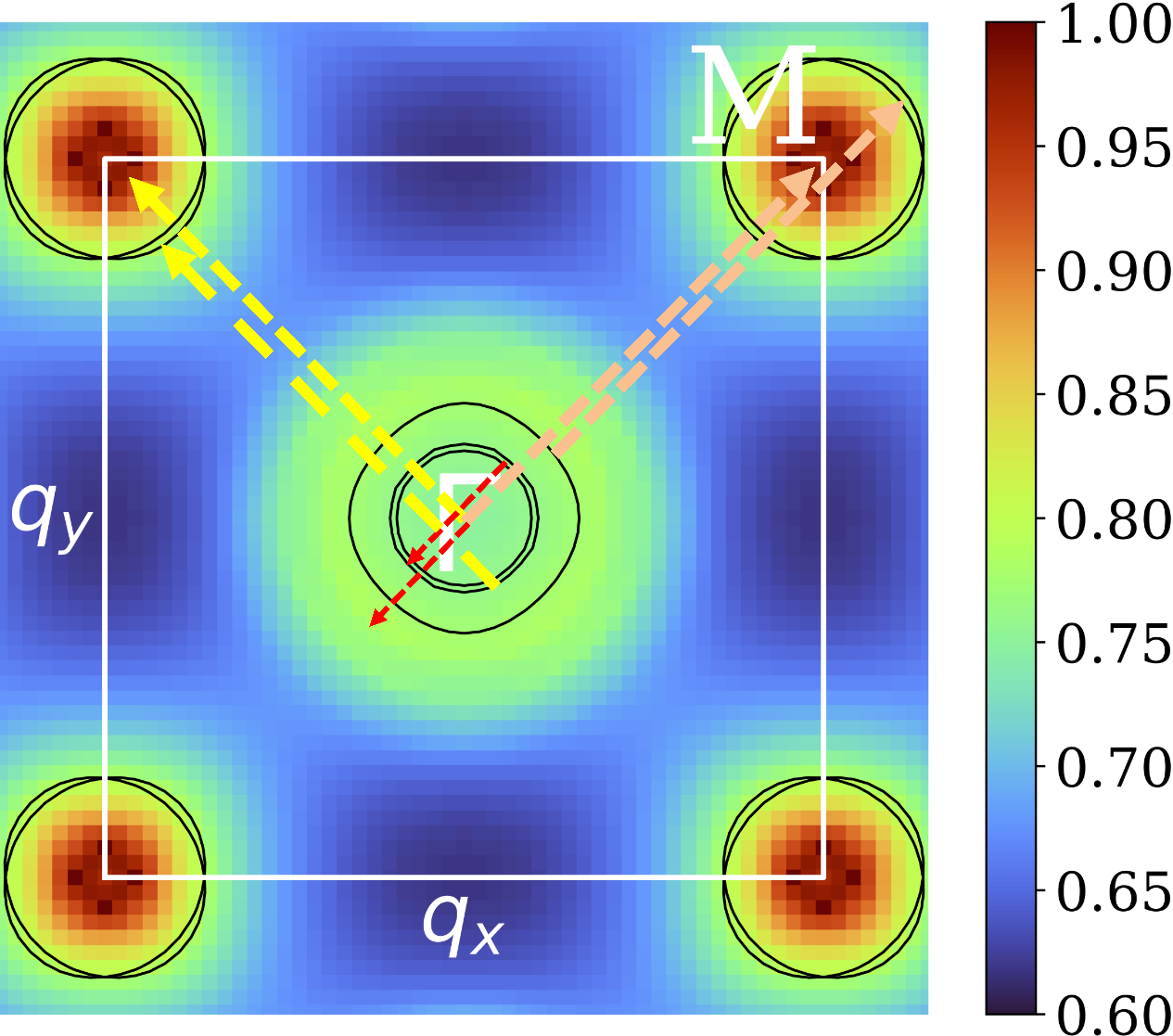}
\caption{Bare susceptibility $\chi^0 ({\bf q})$ on the $q_z = 0$ plane. $\chi^0 ({\bf q})$ is normalized 
by its maximum value. The minimum of $\chi^0 ({\bf q})$ is 0.61 and the maximum of the peak-ring 
circulating $\Gamma$ is 0.78.  Black circles centered at $\Gamma$ and M represent, respectively, 
the cross-sections of hole and electron FS cylinders. Two identical orange (yellow) arrows 
point along the [1,1] ([-1,1]) direction: One arrow denotes wavevector {\bf Q} at which $\chi^0 ({\bf q})$ 
peaks and the other indicates the SF pairing interaction which leads to opposite sign SC gaps on one 
hole and one electron FS pocket at wavevectors {\bf k} and {\bf k} + {\bf Q}, i.e., 
$\Delta_{\bf k} = -\Delta_{{\bf k} + {\bf Q}}$. Similarly, one red arrow denotes wavevector {\bf Q} at which $\chi^0 ({\bf q})$
peaks and the other red arrow denotes the SF pairing interaction between hole pockets H1 and H2 with 
$\Delta_{\bf k} = -\Delta_{{\bf k} + {\bf Q}}$.
}
\label{fig:Lindhard}
\end{figure}

{\it Discussion and conclusions}. To gain a simple picture on the formation of the multiband sign changing 
$d_{xy}$-wave gap structure in ThFeAsN, we depict its bare susceptibility $\chi^0({\bf q})$ on
the $q_z = 0$ plane in Fig.~\ref{fig:Lindhard}. 
There are broad circular peaks at M and also a peak ring circulating $\Gamma$ due to
the inter-FS pocket nesting, as indicated by the orange, yellow and red arrows.
A peak in susceptibility $\chi^{0}({\bf q})$ near wavevector ${\bf q} \approx {\bf Q}$ would lead
to a strong repulsive AF interaction $\Gamma_{s}({\bf k},{\bf k}')$ 
between two electrons having wavevectors ${\bf k}$ and ${\bf k}' = {\bf k}+{\bf Q}$.~\cite{Hirschfeld2011} 
Consequently, an isotropic $s$-wave gap function cannot satisfy
the SC gap equation ($T = 0$ K)
$\Delta_{\bf k} = - \sum_{{\bf k}'}\Gamma_{s}({\bf k},{\bf k}')\frac{\Delta_{{\bf k}'}}{2E_{{\bf k}'}}$.
~\cite{Hirschfeld2011}
However, a solution is possible if the gap function changes sign, $\Delta_{\bf k} = - \Delta_{{\bf k} + {\bf Q}}$.
In the present case, $\chi^0({\bf q})$ peaks near ${\bf Q} = (\pi/a, \pi/a)$.
Thus, due to inter-FS pocket AF coupling indicated by the orange and yellow arrows in Fig. ~\ref{fig:Lindhard},
sign changing $d_{xy}$-wave pairing would occur. Similarly, sign changing $d_{xy}$-wave pairing between
hole pockets H1 and H2 could arise due to inter-FS pocket SF pairing-interaction indicated by the red 
arrows in Fig. ~\ref{fig:Lindhard}.

In conclusion, our fully {\it ab initio} SCDFT calculations show that ThFeAsN is a multiband unconventional superconductor 
with the $B_{2g}$ symmetry SC order parameter and FS sheet-dependent sign $d_{xy}$-wave gap functions. 
The calculated $T_c$ of 22.4 K agrees quite well with the experimental value of 29 K~\cite{Wang2016}. 
We demonstrate that the superconductivity in ThFeAsN is driven by SF pairing interaction (see Fig. 2, Table I and Appendix D).
The predicted nodal $d_{xy}$-wave gap functions are consistent with $\mu$SR, NMR and specific heat experiments.
We present distinct SC properties as quasiparticle DOS spectrum and ultrasonic attenuation coefficient
which can be verified by future experiments. \\

{\it Acknowledgments}. The authors acknowledge the support from the National Science and Technology Council 
(NSTC) and National Center for Theoretical Sciences (NCTS) in Taiwan. The authors also thank the 
National Center for High-performance Computing (NCHC) in Taiwan for the computing time.\\

\label{sec:appendixes}
\appendix
{\it Appendix A: Crystal structure and computational details}.
ThFeAsN has the layered tetragonal ZrCuSiAs-type structure with
space group $P4/nmm$ ($D_{4h}^7$) (No. 129) (Fig.~\ref{fig:crystal-bands}), which consists of Th$_2$N$_2$ and
Fe$_2$As$_2$ bilayers. Its unit cell contains two formula units (f.u.).~\cite{Wang2016,Mao2017}
The experimental lattice constants are $a=4.0414$ \AA$ $
and $c=8.5152$ \AA$ $.~\cite{Mao2017}
The Wyckoff positions of Th, Fe, As and N atoms are 2$c$ (1/4, 1/4, $z_{Th}$), 2$b$ (3/4, 1/4, 1/2),
2$c$ (1/4, 1/4, $z_{As}$), and 2a (3/4, 1/4, 0), respectively, with $z_{Th} = 0.1380$ and $z_{As} = 0.6522$.
Each Fe atom is tetrahedrally coordinated by four As atoms, with the Fe-As bond length
being 2.401 \AA$ $. The two different As-Fe-As bond angles are $114.65^\circ$ and $106.95^\circ$.
The As height $h_{As-Fe}$ is 1.296 \AA$ $. The Th-As bond length is 3.370 \AA.
In the present calculations, the experimental lattice parameters~\cite{Mao2017} mentioned above are used.

The normal state electronic structure is calculated based on the DFT
with the generalized-gradient approximation~\cite{Perdew1996}.
The phonon dispersion and EPC matrix elements are calculated using 
the density functional perturbational theory (DFPT).~\cite{Baroni2001}
The plane wave pseudopotential method is used, with the ultrasoft 
pseudopotentials~\cite{Corso2014} taken from the PSlibrary~\cite{Pslibrary}.
All these {\it ab initio} calculations are carried out using the QUANTUM ESPRESSO
package.~\cite{Giannozzi2009,Giannozzi2017}
The large kinetic-energy cutoffs for wave function expansion and for charge density expansion
used, are 105 Ry and 840 Ry, respectively. Throughout this work, the highly efficient optimized tetrahedron
method for Brillouin zone integration~\cite{Kawamura2014} is adopted.
For the self-consistent electronic structure calculations, the non-$\Gamma$-centered $\bf{k}$-mesh of
16$\times$16$\times$8 is used. For the phonon dispersion and EPC matrix element calculations,
the 8$\times$8$\times$4 $\bf{q}$-mesh with threshold for self-consistency of $10^{-14}$ is exploited.
Note that to avoid the singular contribution ( $\propto 1/q^2$ ) from wavevector $q = 0$ to the screened
Coulomb repulsion~\cite{Kawamura2020}, the 8$\times$8$\times$4 $\bf{q}$-mesh is shifted away from the 
$\Gamma$ point in the BZ, and also along the [001] direction to comply with the tetragonal symmetry
of ThFeAsN.

\begin{table}[b]
\caption{\label{tab:symmetry}%
Irreducible representations (IRREPs) and basis functions for spin singlet even-parity pairing states in
tetragonal crystals with the $D_{4h}$ point group symmetry.~\cite{Annett1990,Yip1993} }
\begin{ruledtabular}
\begin{tabular}{cc}
IRREP  & Basis functions \\ \hline
$A_{1g}$  & $1, (x^2 + y^2), (3z^2-r^2)$  \\
$A_{2g}$  & $x y (x^2 - y^2)$             \\
$B_{1g}$  & $x^2 - y^2$                   \\
$B_{2g}$  & $xy$                          \\
$E_g   $  & $z(x, y); z(x^3, y^3)$        \\
\end{tabular}
\end{ruledtabular}
\end{table}
We use the SCTK code~\cite{Kawamura2020,sctk} to solve the SCDFT gap equation [Eq. (1)]
and to calculate the SC properties of ThFeAsN.
To obtain accurate SC properties, the charge and spin susceptibility, kernel $K$ and
renormalization $Z$~\cite{Kawamura2020} are computed on a dense ${\bf{k}}$-point mesh
of 48$\times$48$\times$24. The charge and spin susceptibilities are calculated using the DFT 
within either the random phase approximation (RPA)~\cite{Gell-Mann1957,Kawamura2020} 
or adiabatic local density approximation (ALDA)~\cite{Zangwill1980,Tsutsumi2020}.
We find that the SC properties calculated with the RPA and ALDA do not differ significantly.
Thus, here we present the calculated SC properties using the ALDA only.
The threshold for self-consistency of $10^{-18}$ is used in the SC gap function calculations. 
The calculated physical quantities on the FS are displayed
using the FermiSurfer program~\cite{Kawamura2019}.

{\it Appendix B: Irreducible representations and basis functions for point group $D_{4h}$.}
ThFeAsN crystalizes in a centrosymmetric tetragonal structure with
the $D_{4h}$ point group symmetry. Table II lists the irreducible
representations (IRREPs) and symmetry-allowed basis functions for
spin-singlet pairing states under $D_{4h}$~\cite{Annett1990,Yip1993}.

{\it Appendix C: Calculation of ultrasonic attenuation coefficient.}
Ultrasonic attenuation coefficients ${\bf \alpha}(T)$ are calculated using the
obtained SC gap function $\Delta_{n\bf{k}}$.
The longitudinal ultrasonic attenuation coefficient is given by~\cite{Schrieffer1983}
\begin{align}
\alpha = & \sum_{nm\bf{k}}g_{mn,\nu}({\bf k},{\bf q})(1 + \frac{\xi_{n{\bf k}}\xi_{m{\bf k}+{\bf q}} - \Delta_{n{\bf k}}\Delta_{m{\bf k}+{\bf q}}}{E_{n{\bf k}}E_{m{\bf k}+{\bf q}}}) \nonumber \\ & (f(E_{n{\bf k}})-f(E_{m{\bf k}+{\bf q}}))\delta(E_{m{\bf k}+{\bf q}}-E_{n{\bf k}}-\omega_{{\bf q}\nu}),
\end{align}
where $\omega_{{\bf q}\nu}$ is the eigenfrequency of phonon mode $\nu$ at ${\bf q}$, $g_{mn,\nu}({\bf k},{\bf q})$
is the EPC matrix element between electronic states ${m{\bf k}}$ and ${n{\bf k}+{\bf q}}$,
and $f(\varepsilon)$ is the Fermi function. 
Here we consider the acoustic phonons only, i.e., assuming $\omega_{q\nu} = v_{\nu}q$.
We also ignore the interband excitations (i.e., $m = n$) 
since phonon frequency and wavenumber are very small.
As a result, we can obtain~\cite{Kawamura2015}
\begin{align}
\alpha = -\sum_{n\bf{k}}g_{nn,\nu}({\bf k},{\bf q})\frac{2\xi_{n{\bf k}}^2}{E_{n{\bf k}}^2} \frac{\partial f(E_{n{\bf k}})}{\partial E_{n{\bf k}} } \delta(\frac{ {\bf v}_{F,n{\bf k}}\cdot {\bf \epsilon}_q}{v_{\nu}} - 1) 
\end{align}
where ${\bf \epsilon}_{\bf q} = {\bf q}/q$ and ${\bf v}_{F,n{\bf k}} = \nabla_{\bf k}\xi_{n{\bf k}}$.
To further simplify the calculations, we make the reasonable assumptions of 
(1) the phonon velocity ($v_{\nu}$) is much smaller than the electron Fermi velocity
($v_F$) (i.e., $v_{\nu} \ll v_F$), (2) $\partial \Delta(\xi_{n\bf{k}})/\partial \xi_{n\bf{k}} \ll 1$
in the vicinity of $E_F$ and $\Delta_{n{\bf k}}/\varepsilon_{n{\bf k}}$ decreases rapidly
with increasing distance from $E_F$
and (3) the EPC of the acoustic phonon with long wave number is insensitive to 
electron band index $n$ and wavevector $\bf{k}$
[i.e., $g_{mn,\nu}(\bf{k},\bf{q})$ can be replaced with its averaged value $g$].
Consequently, the normalized longitudinal attenuation coefficient can be simplified as~\cite{Kawamura2015}
\begin{equation}
\frac{\alpha_S}{\alpha_N}=\frac{\sum_{n{\bf k}}\delta(\xi_{n{\bf k}})\delta(|{\bf v}_{F,n{\bf k}}\cdot {\bf \epsilon}_{{\bf q}}|)2f(|\Delta_{n{\bf k}}|)}{\sum_{n{\bf k}}\delta(\xi_{n{\bf k}})\delta(|{\bf v}_{F,n{\bf k}}\cdot{\bf\epsilon}_{\bf{q}}|)},
\label{eq:ultrasonic}
\end{equation}
which is used in the present calculations. Here $\alpha_N$ is the normal state ultrasonic attenuation coefficient.
Eq.~(\ref{eq:ultrasonic}) includes the line integral on the FS where the Fermi velocity is perpendicular to 
the phonon propagation direction (${\bf v}_{F,n{\bf k}}\cdot {\bf \epsilon}_{{\bf q}} = 0$).
For each line segment, the line integral
\begin{align}
\alpha_{\textbf{k}_1\rightarrow \textbf{k}_2} \equiv
\int_{\textbf{k}_1\rightarrow \textbf{k}_2} \frac{|d \textbf{k}|}{|\nabla_\textbf{k} \xi_{\textbf{k}}| |\nabla_\textbf{k} (\textbf{v}_{\textrm{F},\textbf{k}}\cdot{\bf \epsilon}_{{\bf q}})|} f(|\Delta_\textbf{k}|)
\end{align}
can be computed analytically using the trapezoidal rule, which gives
\begin{align}
C_\textbf{k} \frac{
\ln\left( 1 + \tanh\frac{|\Delta_{\textbf{k}_1}|}{2k_B T}\right) - \ln\left( 1 + \tanh\frac{|\Delta_{\textbf{k}_2}|}{2k_B T}\right)
}{|\Delta_{\textbf{k}_1}|-|\Delta_{\textbf{k}_2}|},
\end{align}
for a nodeless ($\Delta_{\textbf{k}_1} \Delta_{\textbf{k}_2}>0$), and
\begin{align}
C_\textbf{k} \frac{
\ln\left( 1 + \tanh\frac{|\Delta_{\textbf{k}_1}|}{2k_B T}\right) + \ln\left( 1 + \tanh\frac{|\Delta_{\textbf{k}_2}|}{2k_B T}\right)
}{|\Delta_{\textbf{k}_1}|+|\Delta_{\textbf{k}_2}|},
\end{align}
for a nodal ($\Delta_{\textbf{k}_1} \Delta_{\textbf{k}_2}<0$), as well as
\begin{align}
\frac{C_\textbf{k}}{2} \left(1-\tanh\frac{|\Delta_{\textbf{k}_1}+\Delta_{\textbf{k}_2}|}{4k_B T}\right),
\end{align}
for a degenerated $\Delta_{\textbf{k}_1} \approx\Delta_{\textbf{k}_2}$ case.
Here $C_\textbf{k} = |\textbf{k}_2 - \textbf{k}_1|/(|\nabla_\textbf{k} \xi_{\textbf{k}}| |\nabla_\textbf{k} (\textbf{v}_{\textrm{F},\textbf{k}}\cdot{\bf \epsilon}_{{\bf q}})|)$ is a constant for each line segment.
This method enable us to compute stable and accurate ultrasonic attenuation coefficients even for highly anisotropic (non-fully gapped) SC gaps.

\begin{figure*}[!htb]
\includegraphics[width=1.6\columnwidth]{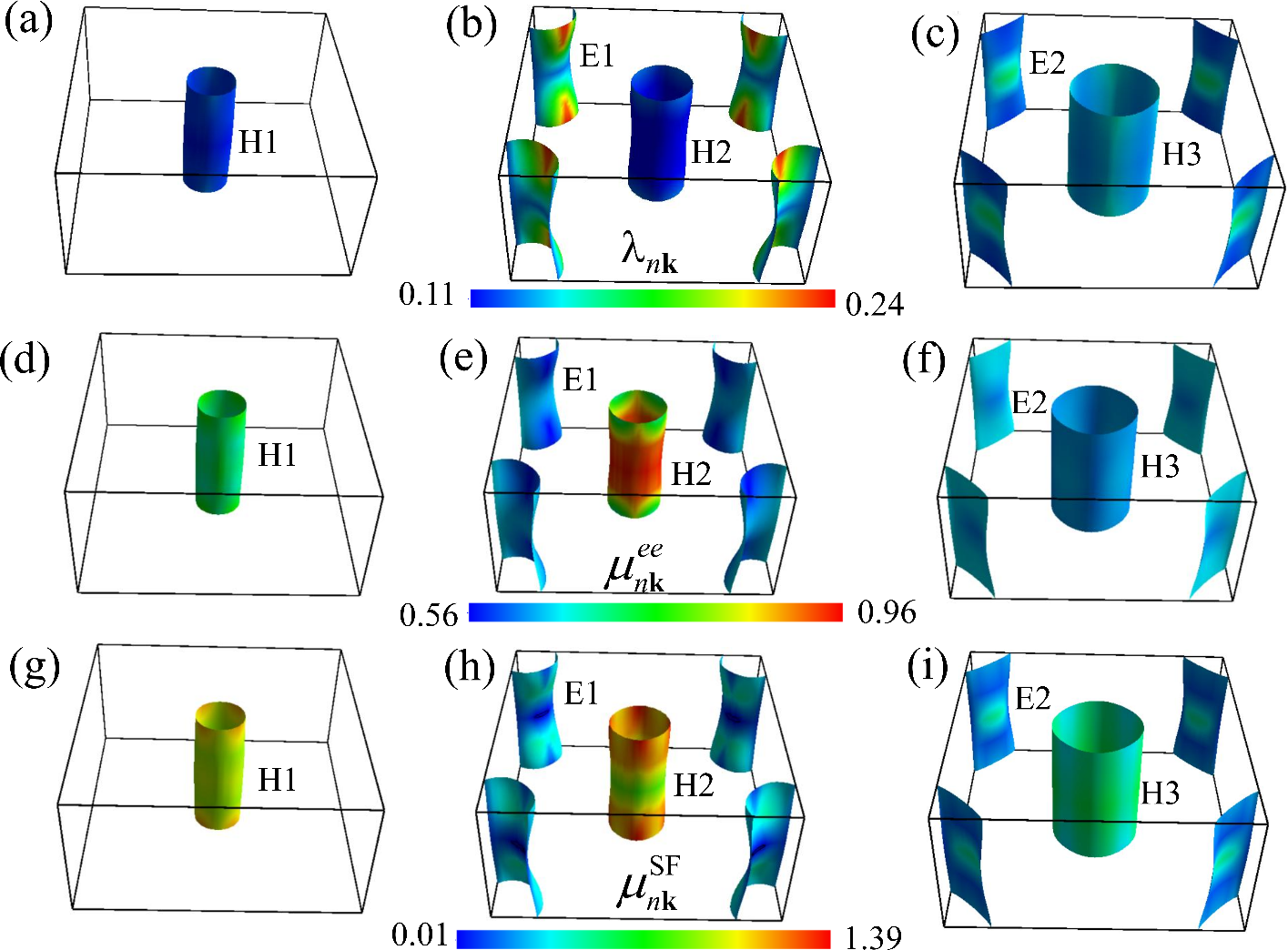}
\caption{Band $n$- and momentum ${\bf k}$-dependent (a,b,c) electron-phonon coupling strength $\lambda_{n\bf{k}}$,
(d,e,f) screened Coulomb repulsion $\mu_{n\bf{k}}^{ee}$, and (g,h,i) spin-fluctuation pair-interaction
$\mu_{n\bf{k}}^{SF}$ on the Fermi surface of ThFeAsN.
}
\label{fig:lambda_FS}
\end{figure*}

{\it Appendix D: Electronic state $n${\bf k}-resolved electron-phonon coupling, screened Coulomb repulsion
and spin fluctuation interaction on the Fermi surface}.
To obtain a detailed understanding of the SC pairing mechanism,
we present the calculated band $n$- and momentum ${\bf k}$-dependent EPC strength~\cite{Giannozzi2017} 
\begin{equation}
\lambda_{n{\bf k}} = \sum_{\nu m}\int_{BZ}\frac{d{\bf q}}{\Omega_{BZ}}\frac{|g_{mn,\nu}({\bf k},{\bf q})|^{2}}{N_F\omega_{{\bf q}\nu}} \delta(\varepsilon_{n{\bf k}}-\varepsilon_F)\delta(\varepsilon_{m{\bf k}+{\bf q}}-\varepsilon_F),
\end{equation}
screened Coulomb repulsion~\cite{Kawamura2020}
\begin{equation}
\mu_{n\bf{k}}^{ee}=\sum_{m{\bf k}'}\delta(\xi_{m{\bf k}'})K_{n{\bf k}m{\bf k}'}^{ee}(\xi_{n{\bf k}},\xi_{m{\bf k}'})
\end{equation}
and SF-induced pairing interaction~\cite{Kawamura2020}
\begin{equation}
\mu_{n\bf{k}}^{SF}=\sum_{m{\bf k}'}\delta(\xi_{m{\bf k}'})K_{n{\bf k}m{\bf k}'}^{SF}(\xi_{n{\bf k}},\xi_{m{\bf k}'})
\end{equation}
in Fig. ~\ref{fig:lambda_FS}.
First, Fig. ~\ref{fig:lambda_FS} indicates that on all the five FS sheets, $\mu_{n{\bf k}}^{ee}$ is always larger 
than  $\lambda_{n{\bf k}}$. Since the signs of the Coulomb repulsion ($\mu_{n{\bf k}}^{ee}$) and phonon-mediated 
attraction ($\lambda_{n{\bf k}}$) are opposite, the screened Coulomb repulsion overwhelms the phonon-mediated 
attraction on all the FS pockets, thus suppressing the weak phonon-induced superconductivity in ThFeAsN (Table I).
Second, Fig. ~\ref{fig:lambda_FS} shows that the $n{\bf k}$-dependent SF-induced pairing
interaction $\mu_{n{\bf k}}^{SF}$ is stronger than both $\mu_{n{\bf k}}^{ee}$ and $\lambda_{n{\bf k}}$
on, at least, all three hole pockets H1-H3. This results in the dominating SF-mediated e-e pairing and hence
unconventional $d_{xy}$-wave  superconductivity in ThFeAsN.


{}	

\end{document}